\documentclass[ reprint,
 amsmath,amssymb,
 aps,
]{revtex4-1}

\usepackage{color}
\usepackage{graphicx}
\usepackage{dcolumn}
\usepackage{bm}% bold math

\newcommand{\be}{\begin{equation}}
\newcommand{\ee}{\end{equation}}
\newcommand{\bea}{\begin{eqnarray}}
\newcommand{\eea}{\end{eqnarray}}

\newcommand{\la}{\langle}
\newcommand{\ra}{\rangle}

\newcommand{\lp}{\left(}
\newcommand{\rp}{\right)}

\renewcommand{\vec}[1]{{\bf #1}}
\renewcommand{\phi}{\varphi}
\renewcommand{\epsilon}{\varepsilon}

\newcommand{\sgn}{\,\text{sgn}\,}

\begin{document}

\preprint{}

\title{Ballistic Guided Electron States in Graphene}
% Graphene}
%Guided `fiber-optic' modes 
%at delta function barrier 
% for 2D Dirac fermions near graphene edge}% Force line breaks with \\
%%\thanks{A footnote to the article title}%

\author{Kamphol Akkaravarawong, Oles Shtanko, Leonid Levitov}
 \affiliation{Physics Department, Massachusetts Institute of Technology, 77 Massachusetts Avenue, Cambridge MA02139}
\date{\today}

\begin{abstract}
Guiding electronic waves in a manner similar to photon transmission in optical fibers is key for developing the electron-optics toolbox. Here we outline a `weak guiding' approach, in which efficient diffraction around disorder results in low-loss, high-finesse electron guiding. We describe an implementation of this scheme for guiding along a narrow-width line gate in gapless and gapped graphene. A simple model for weak guiding, which relies on the Jackiw-Rebbi midgap states, is introduced and solved. The weak-guiding modes are shown to exist for confining potential of either sign and no matter how strong or weak. Modes evolve in a cyclic manner upon varying gate potential, repeatedly sweeping the Dirac gap and becoming dispersionless (flat band) at certain magic values of gate potential. Large mode widths facilitate diffraction around disorder in the core region, enabling exceptionally large mean free paths and long-range ballistic propagation.
%\addLL{List some deliverables: tunable ballistic transport, confined plasmons...}
\end{abstract}

\maketitle

Charge carriers in graphene, because of their steep light-like dispersion resembling that of electromagnetic waves, feature wavelengths that can reach hundreds of nanometers
%$0.1$ micron 
for practical carrier densities~\cite{RMP1}. The quality of graphene samples has been improving rapidly in the past few years, currently yielding mean free paths in the tens of microns, which persist up to room temperature~\cite{wang13,banszerus15}.
%~\cite{mayorov11,rickhaus13,grushina13}{wang13,banszerus15}. 
This unique combination of properties makes graphene an appealing platform for developing quantum electron optics, which aims at controlling electron waves in a fully coherent fashion~\cite{katsnelson2006,cheianov2007,shytov2008,young2009}.
In this vein, it is of great interest to demonstrate guiding of electronic waves in a manner analogous to photons in optical fibers. A number of schemes for electronic waveguides have been proposed, including guiding by a refraction-based confinement, by antidot arrays and by snake states at gate-defined p-n junctions~\cite{pereira2006,zhang2009,hartmann_2010,myoung2011,williams2011,pedersen12,hartmann_2014,rickhaus_2015}. Recent experimental attempts, however,  
were only partially successful, demonstrating the guiding principle but 
stopping short of reaching the anticipated long-range ballistic transport regime~\cite{williams2011,rickhaus_2015}.  

Prior to introducing a new approach, we lay out a few general criteria the electron waveguides must satisfy to achieve high carrier transmission. One requirement is minimizing the number of modes supported by the waveguide. Indeed, since different modes typically propagate at different velocities, multimode waveguides are prone to losses due to carrier scattering between modes as well as signal distortion over long distances. 
It is therefore of interest to focus on the waveguides that support a limited number of modes, preferrably just one mode. Another key requirement is optimizing the guided mode profile in order to reduce scattering by disorder. For designs using a line gate as a waveguide core, most modes can be suppressed by making the gate narrower and/or by tuning the gate potential to the regime of weak confinement. This avoids propagation of most modes and defines a {\it weak guiding} regime for the modes that do propagate. 

\begin{figure} %[h]
\centering\includegraphics[width=1.0\linewidth]{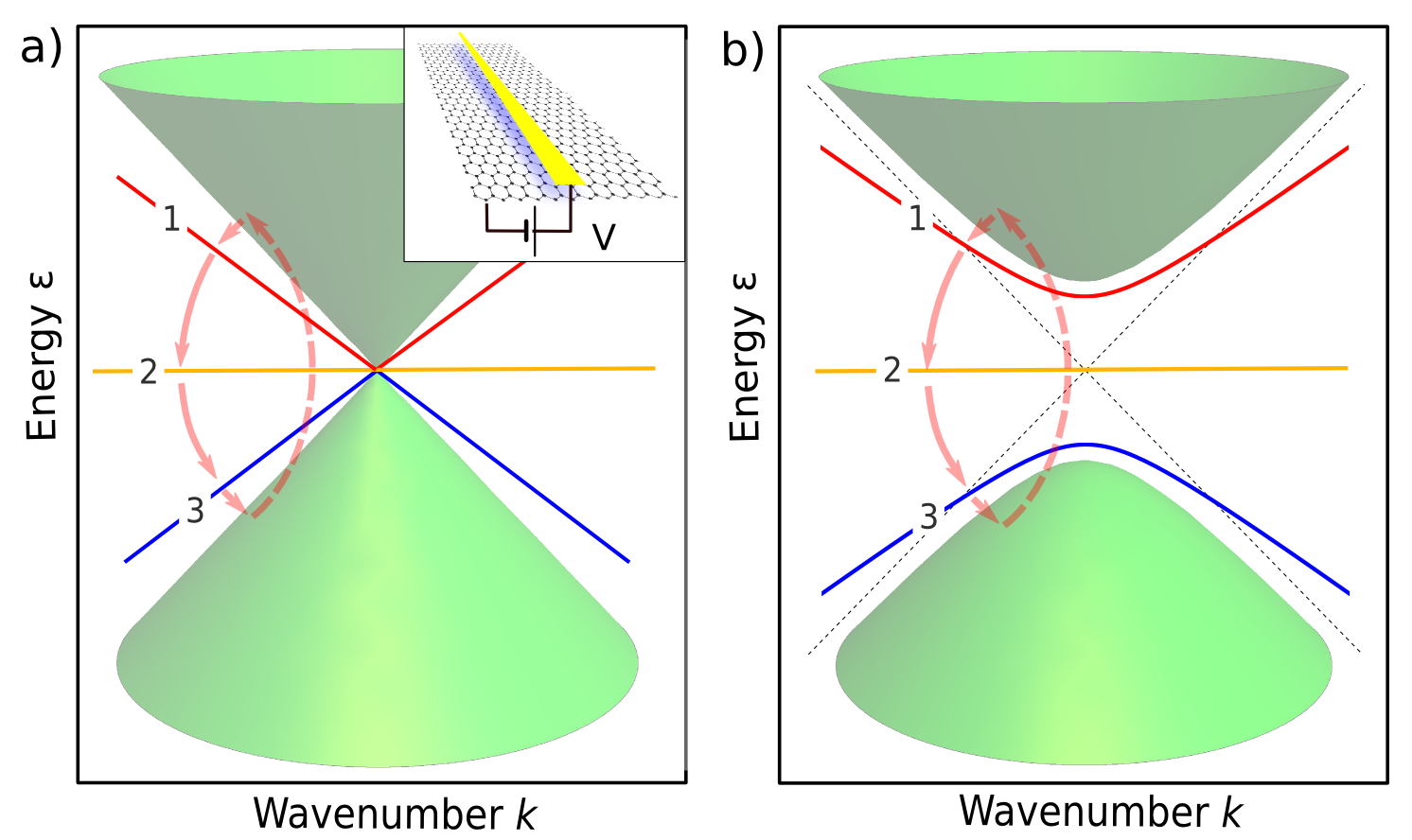}
%{spectrum_MLG_massive.png}
%{out.png}
\caption{Weak guiding mode spectrum for 2D Dirac particles, gapless (a) and gapped (b). The modes, confined by a narrow-width core potential (see inset), propagate as plane waves along the core and decay as evanescent waves into the pristine bulk. Large decay length allows the modes to diffract around disorder in the core region, leading to large mean free paths and long-range ballistic propagation. The weak guiding regime, realized for potential widths smaller than electron wavelengths, $w\lesssim\lambda_\epsilon$, features exactly one guided mode irrespective of the potential strength or sign. Modes evolve in a cyclic manner $...\to1\to2\to3\to1\to...$ as a function of the gate potential strength, see Eq.(\ref{eq:w=uk}). Repeatedly sweeping the gap between Dirac electron and hole continuums (marked by green), modes become dispersionless (flat band) at the `magic' potential values given in Eq.\eqref{eq:flatband}. 
}
\label{fig1}
%\vspace{-5mm}
\end{figure}

Weakly confined modes spread far outside the waveguide core, decaying slowly as evanescent waves in the pristine graphene bulk. A salient feature of this regime is suppression of scattering by disorder in the gated core region, arising because the large width allows such modes to diffract around the disorder at the core. As a result, the disorder scattering rate quickly diminishes at long wavelengths, scaling as a square of the carrier wavenumber. For gapless graphene we find
\be\label{eq:gamma(k)}
\gamma(k)\approx a k^2
,\quad
a=\frac{\alpha}{2\hbar^2 v}\sin\theta_V
\ee 
where $\alpha$ and $\theta_V$ are the disorder strength and the confining potential strength defined in Eq.\eqref{eq:gaussian_fluctuations} and Eq.\eqref{eq:w=uk}, respectively. For realistic carrier wavelength of $100$ nm this predicts mean free paths as large as a few hundred micron (assuming $\alpha$ of an atomic scale). 
%\addLL{add an estimate, comparison to exp, discussion of validity}

Notably, the criteria outlined above are quite familiar in optical fibers, where confinement by a narrow core is known to provide higher performance than refraction-based confinement\cite{gloge1971}. Nearly all optical fibers manufactured at this time can be classified as weakly guiding\cite{buck2004}. Extending this approach to the electronic domain can help achieve control of electron waves at a level comparable to that for light in optical communication systems.

Several approaches to realize weak confinement can be exploited. One option is to use narrow line gates, e.g.  realized with nanowires~\cite{young2009}. This approach, because of its dual-gate tunability, will allow to explore the interesting recurrent behavior of the guided modes which sweep through the Dirac gap in a cyclic manner, as illustrated in Fig.\ref{fig1}. Another approach is to guide carriers along graphene edges where the confining edge potential results from near-edge band bending due to edge doping or electrostatic effects\cite{woessner2015,silvestrov2008}. Being atomically sharp, graphene edges produce a narrow-width potential, providing a natural vehicle for the weak guiding transport~\cite{yacoby_2015}.  
The edge-based approach, while lacking tunbaility, may be appealing because of the ease with which band bending at the graphene edge can be realized. Both approaches lead to weak confinement of the guided waves in the transverse direction, suppressing disorder scattering at the core.

As we will see, the weakly guided modes form robustly, with no set threshold in potential strength. These modes arise for the gate potential (or, edge potential) of either sign, positive or negative, no matter how weak. 
In a complete analogy with optical fibers, the mode lies outside the Dirac continuum (see Fig.\ref{fig1}a,b). For the core potential $V(y)$ of width smaller than the carrier wavelength, $w\ll \lambda_\epsilon=\hbar v/\epsilon$, mode dispersion takes a simple form 
\be\label{eq:w=uk}
\epsilon(k)=\eta \cos(\theta_V)\epsilon_0(k)
,\quad
%\eta=-{\rm sgn\,} ( \Delta\sin\theta_V)
\theta_V=\frac1{\hbar v}\int_{-\infty}^{\infty} V(y)dy
.
\ee
Here $\epsilon_0(k)= \sqrt{v^2 k^2+\Delta^2}$ is the dispersion of gapped Dirac particles in the graphene bulk, 
 $v\approx 10^6\,{\rm m/s}$ is the carrier velocity, and the sign $\eta=\pm 1$ is defined in Eq.\eqref{eq:eta}. The condition $-1\le \cos\theta_V\le 1$ ensures that the mode \eqref{eq:w=uk} is situated outside the Dirac continuum and is thus confined at all $k$. Furthermore, the mode lies within the bulk gap for wavenumbers 
$|k|<\frac{\Delta}{\hbar v}|\tan\theta_V|$.
%$-k_V<k<k_V$, $k_V=\frac{\Delta}{\hbar v}|\tan\theta_V|$ for any gate potential strength. 

Importantly, the narrow-width condition $w\ll \lambda_\epsilon$ for the gate potential is the only validity condition for the result \eqref{eq:w=uk}, which holds for any potential strength, no matter how  weak or strong. 
% The result \eqref{eq:w=uk} holds for any potential strength, which can be either weak or strong. 
The weak guiding regime is therefore robust irrespective of the gate potential strength. 
Accordingly, the phase  $\theta_V$, 
%in the weak guiding regime, described by 
given in Eq.\eqref{eq:eta}, may be order-one or even greater than one in this regime. 

Modes \eqref{eq:w=uk} feature an interesting evolution, sweeping repeatedly the Dirac gap 
upon varying the gate potential. The modes split off the upper (lower) band edge for a weak $V$  of a negative (positive) sign. Upon $V$ increasing, the modes sweep through the Dirac gap, turning into a  {\it perfectly flat band} when potential reaches the value such that $\theta_V=\pi/2$, and then diving into the Dirac continuum at $\theta_V=\pi$. The process repeats in a cyclic fashion at higher $V$, continuing so long as the condition $w\ll \lambda_\epsilon$ holds. This behavior is illustrated in Fig.\ref{fig1}a,b. 

One surprising aspect of the above behavior is that the system features flat bands for certain `magic' potential strength values. Those occur when $\theta_V=\pi/2+\pi m$ with integer $m=0,\pm1,\pm2...$. This corresponds to potential strength such that
\be\label{eq:flatband}
\int_{-\infty}^{\infty} V(y)dy=(\pi/2+\pi m)\hbar v
.
\ee
The flat band behavior, described by Eq.\eqref{eq:flatband}, occurs in a recurrent fashion because modes sweep cyclically through the Dirac gap upon variation of $V$.

Turning to the analysis, we consider electronic states 
in the presence of a line potential $V(y)$ in a graphene monolayer sheet spanning the entire plane.
% (no edges).
% taken to be localized in the interval $-d<x<d$.
%  taken to be a delta function, $V(x)=\lambda\delta(x)$. 
Focusing on one valley, as appropriate for carrier wavelengths larger than graphene lattice constant, we  seek solutions of the Dirac equation $(\epsilon-H)\psi(x,y)=0$ with
\be\label{eq:H_full}
H= v\boldsymbol{\sigma} {\boldsymbol p}+\Delta\sigma_3+V(y)
,
%\epsilon\psi(x,y)=\left( v\boldsymbol{\sigma} {\boldsymbol p}+\Delta\sigma_3+V(y)\right)\psi(x,y)
% ,\quad v\approx 10^6\,{\rm m/s}
\ee
where $\sigma_i$ are Pauli matrices. 
Propagating modes are described by $\psi(x,y)=e^{ikx}\phi_k(y)$ where $k$ is the wavevector component along the line and $\phi_k(y)$ is a two-component (pseudo)spin wavefunction depending on the transverse coordinate. 
Solutions of this equation in the gapless case $\Delta=0$ have been explored numerically and analytically for different forms of $V(y)$~\cite{pereira2006,zhang2009,hartmann_2010,myoung2011,williams2011,hartmann_2014}, however the weak guiding modes and associated long-range ballistic propagation regime have not been identified. 
Here we analyze the narrow-width (or, long-wavelength) regime $wk\ll 1$, where $w$ is the characteristic width of the potential $V(y)$.  We will see that there exist exactly one weak guiding mode per valley. 
In addition, other modes may occur at larger $k$ values 
%which are on the order of the inverse potential width, 
such that $kw\gtrsim 1$~\cite{hartmann_2010,hartmann_2014}. These modes, occurring when potential strength exceeds a certain threshold value, appear to be less robust than the weak guiding modes which exist irrespective of potential being strong or weak.

The above problem can be tackled by a $y$-dependent $SU(2)$ gauge transformation $\psi'=U(y)\psi$ chosen such that it entirely eliminates the potential $V(y)$, 
\be
%\phi(y)\to e^{-i\frac12\theta(y)\sigma_y}\phi(y)
U(y)=e^{i\frac12\theta(y)\sigma_y} 
,\quad
\theta(y)=\frac2{\hbar v}\int_0^y V(y')dy'
.
\ee
Using the identities
\bea \nonumber
% && e^{i\frac12\theta\sigma_y}\sigma_x e^{-i\frac12\theta\sigma_y}
&& U(y)\sigma_x U^{-1}(y)=\sigma_x \cos\theta(y)+\sigma_z \sin\theta(y) ,
\\ \nonumber
% &&e^{i\frac12\theta\sigma_y}\sigma_z e^{-i\frac12\theta\sigma_y}=\sigma_z \cos\theta-\sigma_x \sin\theta
&& U(y)\sigma_z U^{-1}(y)=\sigma_z \cos\theta(y)-\sigma_x \sin\theta(y) ,
\eea
yields a $1D$ Dirac equation $(\epsilon-\tilde H_k)  \tilde\phi_k(y)=0$, where $\tilde H_k=UHU^{-1}$ features a position-dependent mass term: 
\be\label{eq:Dirac_eqn_2}
%\lp\epsilon-H\rp \tilde\phi(y)=0,\quad
%\tilde H=UHU^{-1}
\tilde H_k =-i\hbar v \sigma_y\partial_y+\kappa\sigma^{(k)}_x \cos\theta(y)+\kappa\sigma^{(k)}_z \sin\theta(y)
%e^{i\theta(x)\sigma_x}\sigma_y k e^{-i\theta(x)\sigma_x})
%\tilde\phi(y)
.
\ee
Here we defined, as a shorthand notation, a parameter $\kappa=\epsilon_0(k)$ and a set of $k$-dependent rotated Pauli matrices
\be\label{eq:tilde_sigma}
\sigma^{(k)}_x=\frac{vk}{\kappa}\sigma_x+\frac{\Delta}{\kappa}\sigma_z
,\quad
\sigma^{(k)}_z=\frac{vk}{\kappa}\sigma_z-\frac{\Delta}{\kappa}\sigma_x
.
\ee
The matrices in \eqref{eq:tilde_sigma}, along with $\sigma_y$, obey the standard Pauli algebra. 
% where we simplified the last term with the help of the identity $e^{i\theta\sigma_x}\sigma_y e^{-i\theta\sigma_x}=\sigma_y \cos(2\theta)-\sigma_z \sin(2\theta)$. 
The Dirac Hamiltonian $\tilde H_k$ 
%in Eq.\eqref{eq:Dirac_eqn_2} 
features a mass term, defined by a matrix that rotates as a function of $y$, but no external potential.

An explicit analytic solution can be constructed for narrow-width potentials $V(y)$ as follows. 
Focusing on the long-wavelength modes, $kw\ll1$, where $w$ is the potential width, we approximate 
% Eq.\eqref{eq:Dirac_eqn_2} with 
$\theta(y)\approx \theta_V\sgn (y)$ with the $\theta_V$ value 
given in Eq.\eqref{eq:w=uk}. Replacing $\cos(\theta(y))$ by a constant $\cos\theta_V$ and $\sin(\theta(y))$ by $\sin\theta_V\sgn(y)$, gives a 1D Dirac Hamiltonian with a mass kink $m(y)=(\kappa\sin \theta_V)\sgn(y)$: 
\be\label{eq:Dirac_eqn_JR}
%\lp \epsilon-\tilde H\rp \tilde\phi(y)=0
%,\quad 
\tilde H_k=-i\hbar v \sigma_y\partial_y+\kappa_V \sigma^{(k)}_x + m(y)\sigma^{(k)}_z
%\quad \tilde k=k\cos u, \ m(y)=(k\sin u){\rm sgn\,}(y)
\ee
where we defined $\kappa_V =\kappa\cos \theta_V$. %, $m(y)=(\kappa\sin \theta_V)\sgn(y)$. 

In what follows we will assume the potential width $w$ to be smaller that the gap-induced length $\ell=\hbar v/\Delta$. For $\Delta$ of a few tens of meV (a realistic value for G/hBN heterostructures), the length $\ell$ is in the tens of nm. 
%\addLL{[discuss validity, generality, robustness due to JR results, add justification?]}
In the regime $w\ll \ell$, electronic guided modes can be constructed directly and explicitly in terms of the Jackiw-Rebbi midgap states~\cite{jackiw_1976} confined by the $m(y)$ step and propagating along it as plane waves. This is achieved in two steps. We first suppress the term $\kappa_V \sigma^{(k)}_x$ in \eqref{eq:Dirac_eqn_JR} and construct an eigenstate with $\epsilon=0$ by solving
\be
i\hbar v \sigma_y\partial_y \tilde\phi(y) = m(y)\sigma^{(k)}_z \tilde\phi(y)
\ee
giving $\tilde\phi_0(y)=\exp(\sigma^{(k)}_x\int_0^y m(y')dy')\tilde\phi(0)$. A normalizable eigenstate is obtained by 
%fixing $\tilde\phi(0)$ equal 
picking a suitable eigenstate $\sigma^{(k)}_x|\tilde\phi(0)\ra=\eta|\tilde\phi(0)\ra$, with $\eta=\pm 1$ given by
\be\label{eq:eta}
\eta=-\sgn( m_{y>0}-m_{y<0})=-\sgn(\sin\theta_V)
.
\ee 
Eigenstates with $\epsilon\ne 0$  can be constructed as products of $\tilde\phi_0(y)$ and plane-wave factors: $\tilde\psi_k(x,y)=e^{ikx}\tilde\phi_0(y)$. 
%Guided-wave states for such a Hamiltonian are described as products of the zero-mode state found for $\tilde k=0$ and a plane wave factor $e^{ikx}$. The energies of these states are simply $\epsilon=\pm\kappa_V $ with the sign given by ${\rm sgn\,} ( m(0+)-m(0-))$. 
This gives the dispersion $\epsilon(k)=\eta\kappa_V$, leading to Eq.\eqref{eq:w=uk}.
% with the sign given by ${\rm sgn\,}(\sin \theta_V)$. 
Since $|\cos\theta_V|<1$, 
%for each $k$ 
the energies of these states lie outside the bulk continuum 
$|\epsilon|>\epsilon_0(k)$, which ensures decoupling from the bulk states and confinement to the region near the $y=0$ line. 
%This behavior resembles refraction-based confinement of light in fiber optics. 
The connection with the theory of zero modes~\cite{jackiw_1976}  indicates the robustness of such confinement. 

With the sign $\eta$ chosen as above, the mode is positioned slightly below the upper Dirac band $\epsilon=\epsilon_0(k)$ for a weak negative potential, and slightly above the lower Dirac band $\epsilon=-\epsilon_0(k)$ for a weak repulsive potential, as expected from perturbation theory. 
%However, since our results are applicable for both strong and weak gate potential $V(y)$, 
At a strong potential, however, the sign $\eta$ and the mode energy feature an interesting oscillatory behavior as a function of the potential strength, periodic in $\theta_V$ with a period $\pi$. 

%
%[{\bf Discuss features: phase $\theta_V$ is not necessarily small in the weak guiding regime, reduce duplicity with discussion after Eq.1}]

Next, we apply the above results to estimate the effect of disorder scattering and demonstrate that it is suppressed at small $\epsilon$. We will assume that scattering occurs predominantly within the core region under the line gate and model it by a fluctuating confining potential, % strength,
\be\label{eq:V+dV}
V(x,y)=\lp V_0+\delta u(x)\rp\delta(y)
.
%,\quad
%\la \delta u(x) \delta u(x')\ra =\alpha \delta(x-x')
\ee
Writing the Green's function $G=1/(i\epsilon-H)$, where $H$ is given in Eq.\eqref{eq:H_full} with the potential \eqref{eq:V+dV}, 
as a formal series expansion, we have
\be
G=G_0+G_0(V+\delta V)G_0+G_0(V+\delta V)G_0(V+\delta V)G_0+...
\ee
where $G_0=1/(i\epsilon-v\boldsymbol{\sigma} {\boldsymbol p}-\Delta\sigma_3)$. 
In averaging $G$ over disorder $\delta u(x)$ we use the Gaussian model, i.e. account only for the pair correlators 
\be\label{eq:gaussian_fluctuations}
\la \delta u(x) \delta u(x')\ra =\alpha \delta(x-x')
.
\ee
In a non-crossing approximation, this gives Dyson's equation $\la G\ra=1/(i\epsilon-H-\Sigma)$ with the self-energy operator
\be\label{eq:sigma(E,y)}
\Sigma(\epsilon,y)=\delta(y)S(\epsilon)
,\quad
S(\epsilon)=\alpha\int\frac{dk}{2\pi}G(\epsilon,k,y,y')_{y=y'=0}
.
\ee
%express the result through a suitable self-energy operator 
%\be
%\la G\ra=G_0+G_0(V+\Sigma)G_0+G_0(V+\Sigma)G_0(V+\Sigma)G_0+...
%\ee
%where
%\be\label{eq:sigma(E,y)}
%\Sigma(\epsilon,y)=\alpha\delta(y)\int\frac{dk}{2\pi}G(\epsilon,k,y,y')_{y=y'=0}
%.
%\ee
Crucially, the quantity $\Sigma(\epsilon,y)$ spatial dependence is the same as that of the gate potential $V(y)\approx V_0\delta(y)$. We can therefore incorporate this term in the above solution in a seamless fashion by using a matrix gauge transformation to eliminate the entire effective potential $\lp V_0+S(\epsilon)\rp\delta(y)$ and replace it with a position-dependent mass term. 
% \addLL{OUTSTANDING QUESTION: what about the $\sigma^{(k)}_x$ structure for $V_0\ne 0$?} 

We will illustrate this approach for the gapless case, $\Delta=0$. First, we replace the Greens function in Eq.\eqref{eq:sigma(E,y)} by the bare quantity $G_0$ (a good approximation in the weak perturbation limit, i.e. small $V_0$). Integrating over $k$ and continuing $i\epsilon\to\epsilon$ gives
% in Eq.\eqref{eq:sigma(E,y)} gives
\be\label{eq:S(E)}
S(\epsilon)=-\frac{\alpha\epsilon}{2\pi\hbar^2 v^2}\ln\frac{\epsilon_0}{\delta-i\epsilon}
\ee
where $\epsilon_0$ is a UV regularization parameter introduced to control the divergence of the integral over $k$ and $\delta$ is a positive infinitesimal number. The quantity $S(\epsilon)$ is complex-valued, and vanishes at $\epsilon\to 0$. Therefore, after replacing $V_0$ in Eq.\eqref{eq:w=uk} with $V_0+S(\epsilon)$, we obtain a dispersion relation which accounts for disorder scattering,
\be\label{eq:epsilon_S}
\epsilon=\eta\epsilon_0(k)\cos((V_0+S(\epsilon))/\hbar v).
\ee
% with an imaginary part describing decay due to scattering by disorder. 
Since $S(\epsilon)$ vanishes at $\epsilon\to 0$, we can Taylor-expand the above expression to obtain the decay rate. The latter, 
% The decay rate, 
given by the imaginary part of the mode frequency, $\epsilon=\epsilon'+i\gamma/2$, vanishes at small $k$ as $k^2$ as given in Eq.\eqref{eq:gamma(k)}. 
% \addLL{comment that $S(\epsilon)$ vanishing at $\epsilon\to 0$ vindicates treating it as a perturbation.}

% also predicts logarithmic corrections to mode velocity which can become substantial for potential strength satisfying the flat band condition, $\cos(V_0/\hbar v)=0$, Eq.\eqref{eq:flatband}.
% $V_0=(\pi/2+\pi m)\hbar v$.
% in the long wavelength limit. 

Next, we show that our result for self-energy $S(\epsilon)$ preserves its form (up to a numerical prefactor) for an any $V_0$. This is done by evaluating the Greens function $G=1/(i\epsilon-H)$. We first assume $\Delta\ne0$ but eventually focus on the $\Delta=0$ case. The analysis is most transparent in a mixed position-momentum representation
\be\label{eq:Gkyy'_mixed}
G_{\alpha\beta}(\vec r,\vec r')=\sum_k e^{ik(x-x')}\la\alpha| U(y)G_{k}(y,y')U^{-1}(y')|\beta\ra
.
\ee
Here $\vec r=(x,y)$ and for each $k$ we define the quantity 
\be 
G_{k}(y,y')=\la y'|\frac1{i\epsilon-\tilde H_k}|y\ra
\ee 
which is a $2\times 2$ matrix in (pseudo)spin variables (here $\tilde H_k$ is given in Eq.\eqref{eq:Dirac_eqn_JR}). Inverting the Dirac operator with a position-dependent mass term can be performed with a trick which links
%reduces the evaluation of 
the Green's function for $\tilde H_k$ to that for a suitably defined auxiliary nonrelativistic Schr\"odinger Hamiltonian. To that end we consider an identity
%will be useful in studying transport and compressibility of the system. The analysis can be performed with the help of the identity
%\addLL{[add $U$ and $U^{-1}$; comment that $G_{\epsilon,k}(y,y')$ is an operator in pseudospin?]}
\be\label{eq:Gepsilon,k}
G_{k}(y,y')=\la y'|\frac1{i\epsilon-\tilde H_k}|y\ra
=\la y'|\frac{-i\epsilon-\tilde H_k}{\epsilon^2+\tilde H_k^2}|y\ra
.
\ee 
% where an infinitesimal imaginary part must be incorporated in the energy as $\epsilon\to\epsilon+i0$. 
The operator $\tilde H_k^2$ is nothing but the one-dimensional Schr\"odinger operator with a delta function potential:
% From the identity
\be\label{eq:H^2}
\tilde H_k^2=-\hbar^2 v^2\partial_y^2+\kappa^2+\lambda\kappa\delta(y)\sigma^{(k)}_x
,\quad
\lambda=2\hbar v\sin\theta_V
,
\ee
where we used the relation $\kappa_V ^2+m^2=\kappa^2$.  Separating the delta function term in Eq.\eqref{eq:H^2} and using  the T-matrix approach, we evaluate the inverse of $\epsilon^2+\tilde H_k^2$ as
\be\label{eq:1/(e^2-h^2)}
\frac1{\epsilon^2+\tilde H_k^2}=D_k
-D_k T_{\epsilon,k}|0\ra\la 0| D_k
,\quad
D_k=\frac1{\epsilon^2+\tilde H_{0,k}^2}
.
\ee
Here $\tilde H_{0,k}^2=-\hbar^2 v^2\partial_y^2+\kappa^2$ is the operator in Eq.\eqref{eq:H^2} without the last term, $|0\ra$ is the $y=0$ position eigenstate, and the T-matrix equals
%the T-matrox is defined as 
\be\label{eq:Tek}
T_{\epsilon,k}=\frac{\lambda\kappa\sigma^{(k)}_x}{1+\lambda\kappa\sigma^{(k)}_x\la 0|D_k|0\ra
%\frac{\lambda\sigma^{(k)}_x}{2\pi\hbar}\int\frac{dp_y}{\epsilon^2+\kappa^2+v^2p_y^2}
}
=\frac{\lambda\kappa\sigma^{(k)}_x}{1+\frac{\lambda\kappa\sigma^{(k)}_x}{2\hbar v\sqrt{\kappa^2+\epsilon^2}}}
.
\ee
% must be incorporated in the energy as $\epsilon\to\epsilon+i0$
Mode dispersion is obtained  from the pole 
%of $T_{\epsilon,k}$
in Eq.\eqref{eq:Tek} by continuing to real frequencies and incorporating an infinitesimal imaginary part as required by the causality condition, $i\epsilon\to\epsilon+i0$. 
This gives $\sqrt{\kappa^2-\epsilon^2}=-\kappa\sin\theta_V\sigma^{(k)}_x$.
This relation, after simple algebra, yields the dispersion identical to that in Eq.\eqref{eq:w=uk}.

Using the above results we proceed to evaluate the equal-point Greens function. Setting $x=x'$, $y=y'$, $\Delta=0$ in Eq.\eqref{eq:Gkyy'_mixed},
%(from now on we set $\Delta=0$)
%%spatially-resolved density of states. This quantity  can be expressed through the equal-point Greens function as 
%\be
%\sum_k G_{\epsilon,k}(y,y)=\frac1{2\pi v}\sum_k\frac{-i\epsilon-kv\sin\theta_V\sigma_x}{(k^2v^2+\epsilon^2)^{1/2}+kv\sin\theta_V\sigma_x}
%%N(\epsilon,y)=-\frac1{\pi}{\rm Im}\,\sum_k \Tr G_{\epsilon,k}(y,y)
%% \la y|\frac1{\epsilon-H+i0}|y\ra
%\ee 
%(from now on we set $\Delta=0$). 
% \addLL{LL: OUTSTANDING QUESTION: Check the numerator} 
and estimating the integral over $k$ we arrive at an expression which is identical to Eq.\eqref{eq:S(E)}, behaving at low energies as $\epsilon\ln\frac{\epsilon_0}{\delta-i\epsilon}$ up to a numerical prefactor. The above analysis of Dyson equation then yields Eq.\eqref{eq:epsilon_S}, 
vindicating our conclusion about the disorder scattering rate vanishing as $k^2$ at long wavelengths. 
% due to weakly-guided mode diffraction around disorder in the core region. 

Another interesting question the result in Eq.\eqref{eq:epsilon_S} helps to address is the impact of scattering on the flat band behavior, predicted when $\cos(V_0/\hbar v)=0$, see Eq.\eqref{eq:flatband}. Setting $V_0=(\pi/2+\pi m)\hbar v$ we see that, within the approximations made, the non-dispersing bands with energy $\epsilon=0$ are robust in the presence of disorder scattering.

Summing up, weak guiding of electron waves can be realized in graphene 
%gated by 
with narrow-width line gates. 
%Weakly guided modes, due to their large width, can efficiently diffract around disorder in the gated core and thus provide 
The large width of weakly guided modes allows them to efficiently diffract around disorder in the gated core, yielding 
exceptionally long mean free paths.  Estimates for realistic parameter values predict mean free paths as long as 100 microns. Demonstrating weak guiding will help pave the way to scalable networks of electron waveguides in which electron waves are manipulated with precision and control 
%similar to 
matching that of
%for photons in 
optoelectronic circuits. 

We acknowledge support of the Center for Integrated Quantum Materials
(CIQM) under NSF award 1231319 (OS and LL) and partial support by the U.S. Army Research
Laboratory and the U.S. Army Research Office
through the Institute for Soldier Nanotechnologies, under
contract number W911NF-13-D-0001 (LL).

% \vspace{-8mm}

%\section{Appendix}

\end{document}